\documentclass{webofc}
\usepackage[varg]{txfonts}   
\begin{document}
\title{The muon g-2: retrospective and future}
%
%
\author{A.~E.~Dorokhov\inst{1,2,3}\fnsep\thanks{\email{dorokhov@theor.jinr.ru}} \and
        A.~E.~Radzhabov\inst{4}\fnsep\thanks{\email{aradzh@icc.ru}} \and
        A.~S.~Zhevlakov\inst{3,4}\fnsep\thanks{\email{zhevlakov1@gmail.com}}
}

\institute{ Bogoliubov Laboratory of Theoretical Physics, JINR, 141980 Dubna, Russia\label{addrJINR}
          \and
          N.N.Bogoliubov Institute of Theoretical Problems of Microworld, M.V.Lomonosov Moscow State University, Moscow 119991, Russia\label{addrMSU}
          \and
          Department of Physics, Tomsk State University, Lenin ave. 36, 634050 Tomsk, Russia
          \and
          Institute for System Dynamics and Control Theory SB RAS, 664033 Irkutsk, Russia \label{addrIDSTU}
}

\abstract{%
 Soon, new experiments at FNAL and J-PARC will measure the muon anomalous magnetic moments with better accuracy than before. From theoretical side, the uncertainty of the standard model prediction is dominated by the hadronic contributions. Current status of the experimental data and theoretical calculations are briefly discussed.
}
\maketitle
\section{Introduction}
\label{intro}
Cosmology tell us that about $95\%$ of matter is not detected by modern measurements. We think that the dark matter surround us, however, we don't see it. There are two strategies to search for the physics beyond the standard matter: high energy and low energy experiments. In the first case, due to high energy we attempt to excite the heavy degrees of freedom. There are no firm evidences on the deviation of measured cross sections from the predictions of the standard model. In the case of the low energy experiments, it is possible to reach very high precision of the measured quantities because of huge statistics. And within this kind of experiments there are some rough redges of the standard model. The most famous deviation is observed for the muon anomalous magnetic moment and it remains stable for many years.

\begin{figure}[h]
\begin{center}
\includegraphics[height=6cm]{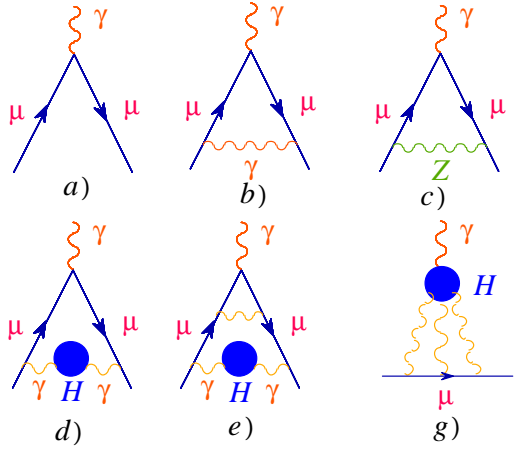}%
\end{center}
\caption{Representative diagrams for the SM contributions to $a_\mu$.
Here, $H$ is for the hadronic block.}
\label{SM}
\end{figure}

The anomalous magnetic moment (AMM) of charged leptons ($l=e,\mu ,\tau $) is
defined as%
\begin{equation}
a_{l}=\frac{g_{l}-2}{2},  \label{a}
\end{equation}%
with the gyromagnetic ratio $g_{l}$ of the lepton magnetic moment to its
spin, in Bohr magneton units. The Dirac equation $g=2$ predicts for a free point-like fermion with spin $1/2$ and thus there is no anomaly at tree level (Fig. \ref{SM}a).
However, deviations appear when taking into account the interactions leading to fermion substructure and
thus to nonzero $a_{l}$. In the standard model it appears from the radiative corrections to the tree fermion-photon vertex (Fig. \ref{SM})
 due to the coupling
of the lepton spin to virtual fields, which in the SM are induced by QED, weak and
strong (hadronic) interactions (Fig. ~\ref{SM})%
\begin{equation}
a^{\mathrm{SM}}=a^{\mathrm{QED}}+a^{\mathrm{weak}}+a^{\mathrm{hadr}}.
\label{aSM}
\end{equation}

\section{The electron $g-2$ and the fine coupling constant}
\label{sec-1}

The electron AMM was experimentally discovered by Kush and Foley from Columbia University, New York (USA) in 1947 \cite{Kush:1947,Kush:1948} with the result \cite{Schwinger:1948iu}
\begin{equation}
a_{e}^{\mathrm{Columbia}}=0.00118\pm0.00003.  \label{aeKushFoley}
\end{equation}

This was immediately confirmed by Schwinger in the framework of quantum electrodynamics as a result of the radiative correction (Fig. \ref{SM}b)
\begin{equation}
a_{e}^{\mathrm{Schwinger}}=\frac{\alpha}{2\pi}=0.001161.  \label{aeSch}
\end{equation}

Since that time, enormous progress has been reached in experiment and theory. The latest measurement by the Gabrielse' group from Harvard university (USA) provides the result with fantastic accuracy \cite{Hanneke:2008tm}
\begin{equation}
a_{e}^{\mathrm{Harvard}}=1~159~652~180.73~(0.28)\times 10^{-12}\quad \lbrack
0.24~\mathrm{ppb}].  \label{aeHarv}
\end{equation}

Within the standard mode (\ref{SM}), the electron AMM is almost completely dominated by QED contribution calculated with 5-loop accuracy
\begin{equation}
a_{l}^{\mathrm{QED}}=\sum_{n=1}^5 C_{2n}^l\big(\frac{\alpha}{\pi}\big)^n+.... , \label{alQED}
\end{equation}
where the value of coefficients are given in Table~\ref{tab-cQED}.

\begin{table}
\caption{The coefficients in the QED contribution to the lepton AMM (\ref{alQED}). The first three coefficients are known analytically. The errors in $C_{4,6}^\mu$ are due to the experimental uncertainties in $m_\mu/m_e$ ratio. The errors in $C_{8,10}^l$ are due to numerical simulations.}
\label{tab-cQED}       
\centering
\begin{tabular}{clll}
\hline
        & l=e & l=$\mu$      & source\\ \hline
$C_{2}^l$ & 0.5 & 0.5& Schwinger\cite{Schwinger:1948iu} \\
$C_{4}^l$ & -0.328~478~444~00... & 0.765~857~425(17)& Laporta, Remiddi \cite{Laporta:1996mq} \\
$C_{6}^l$ & 1.181~234~017... & 24.050~509~96(32)& Laporta, Remiddi \cite{Laporta:1996mq}\\
$C_{8}^l$ & -1.912~06(84) & 130.879~6(63)& Kinoshita, Nio, et.al. \cite{Aoyama:2012wj}\\
$C_{10}^l$ & 7.791(336) & 753.29(1.04)& Kinoshita, Nio, et.al. \cite{Aoyama:2012wj}\\ \hline
\end{tabular}
\end{table}

The result (\ref{alQED}) allows to determine the fine structure constant $\alpha $
with the extraordinary precision \cite{Aoyama:2012wj,KINOSHITA:2014uza}%
\begin{equation}
\alpha ^{-1}(a_e)=137.035~999~1570~(29)(27)(18)(331)\quad \lbrack 0.25~\mathrm{ppb}],
\label{alpha_QED}
\end{equation}%
where the first two uncertainties are due to errors in $C_{8,10}^l$, the third one is uncertainty from hadronic and weak corrections, and the last one is due to experimental error in the measurement of $a_{e}^{\mathrm{Harvard}}$.
This determination became possible after the complete QED contribution to the
electron AMM up to tenth order in the coupling constant were achieved
numerically by the Prof. T. Kinoshita group \cite{Aoyama:2012wj} (for recent
review see \cite{KINOSHITA:2014uza}).

In 2010, the direct determination of the fine coupling constant became possible from measurement of the ratio $\hbar/m_{Rb}$ \cite{Bouchendira:2010es}
\begin{equation}
\alpha ^{-1}(Rb)=137.035~999~049~(90)\quad \lbrack 0.66~\mathrm{ppb}].
\label{alpha_QED_Rb}
\end{equation}
With this $\alpha$, the the SM prediction for the electron AMM becomes
\begin{equation}
a_{e}^{\mathrm{SM,Rb}}=1~159~652~181.643~(764)\times 10^{-12}\quad \lbrack
0.24~\mathrm{ppb}].  \label{aeRb}
\end{equation}
Both determination of the electron AMM (\ref{aeHarv}) and (\ref{aeRb}) are consistent within the errors.
\begin{equation}
a_{e}^{\mathrm{exp}}-a_{e}^{\mathrm{SM,Rb}}=-0.91~(81)\times 10^{-12}.  \label{DeltAae}
\end{equation}
Thus the experimental and SM results for the electron AMM are in perfect agreement.

\section{The muon $g-2$: experiment vs standard model. Electroweak contributions.}
\label{sec-2}

In 2006, there were published the results obtained by the E821 collaboration at Brookhaven National Laboratory \cite{Bennett:2006fi} on  measurements of the muon AMM$a_\mu$
\begin{equation}
a_{\mu }^{\mathrm{BNL}}=116~592~08.0~(6.3)\times 10^{-10}\quad \lbrack 0.54~%
\mathrm{ppm}].  \label{amuBNL}
\end{equation}
Later on, this value was corrected \cite{Mohr:2012tt,Agashe:2014kda} for a
small shift in the ratio of the magnetic moments of the muon and the proton as
\begin{equation}
a_{\mu}^{\mathrm{BNL,CODATA}}=116~592~09.1~(6.3)\times10^{-10}.
\label{amuBNL2}%
\end{equation}

It is well-known that the effect of the second-order contribution, due to exchange by the particle with mass $M$, to the AMM of the lepton with mass $m_{l}$
 is proportional to $a_{l}\propto(m_{l}/M)^{2}$.
It means, that sensitivity for the muon to the interaction with scale $M$ is by factor
$m_\mu^2/m_e^2 \propto 40000$ higher than for the electron. This fact compensates a less experimental
accuracy of the muon AMM  measurements (\ref{amuBNL}) relatively to the electron one (\ref{aeHarv}), and make the study of the muon AMM more
perspective in search for new physics.

Another exciting point  is that soon the new data on the muon AMM will be available
from experiments
proposed at Fermilab (USA) \cite{Venanzoni:2012sq} and J-PARC (Japan) \cite%
{Saito:2012zz}. These experiments plan to reduce the present experimental
error by factor 4, to a precision of $0.14$ ppm.

In SM, the contributions to the muon AMM from QED (Fig.~\ref{SM}b) \cite{Aoyama:2012wk}
 and weak (Fig.~\ref{SM}c) \cite{Czarnecki:2002nt,Gnendiger:2013pva} interactions (Fig.~\ref{SM}c) are known with high accuracy
\begin{eqnarray}
a_{\mu }^{\mathrm{QED,Rb}}=11~658~471.8951~(0.0080)\times 10^{-10},\\
a_{\mu }^{\mathrm{QED,a_e}}=11~658~471.8846~(0.0037)\times 10^{-10},\label{aMuQED}\\
a_{\mu }^{\mathrm{weak}}=15.36~\left( 0.10\right) \times 10^{-10},
\label{aWeak}
\end{eqnarray}%
The most important feature of new estimate for the weak sector, that significantly increases the theoretical
precision, is to use precise Higgs-boson mass value measured at LHC. The remaining theory error comes from unknown three-loop contributions and dominantly from light hadronic uncertainties in the second-order
electroweak diagrams with quark triangle loops. The accuracy of these calculations is enough for any planed experiments in new future.

Subtracting from the experimental result the well-defined contributions from QED and weak interactions one gets
\begin{equation}
a_{\mu }^{\mathrm{BNL}}-a_{\mu }^{\mathrm{QED,Rb}}-a_{\mu }^{\mathrm{weak}}=721.65~\left( 6.38\right) \times 10^{-10},
\label{DaEWeak}
\end{equation}
where the error is only due to the experiment. We can treat this number as an experimental result for the rest contributions, i.e. of the strong interaction of SM and of the hypothetical interactions beyond SM.

\section{Hadronic contributions to the muon $g-2$. Vacuum polarization effect.}
\label{sec-3}

Strong (hadronic) interaction produces relatively small contributions to $a_\mu$, however they are known with an accuracy comparable to the
experimental uncertainty in (\ref{amuBNL}). In the leading in $\alpha$ orders, these contributions can be
separated into three terms%
\begin{equation}
a_{\mu }^{\mathrm{hadr}}=a_{\mu }^{\mathrm{HVP,LO}}+a_{\mu }^{\mathrm{HVP,HO}%
}+a_{\mu }^{\mathrm{HLbL}}.  \label{aStrong}
\end{equation}%
In (\ref{aStrong}), $a_{\mu }^{\mathrm{HVP}}$ is the leading in $\alpha $ contribution due to the hadron vacuum
polarization (HVP) effect in the internal photon propagator of the one-loop
diagram (Fig.~\ref{SM}d), $a_{\mu }^{\mathrm{ho}}$ is the next-to-leading
order contribution related to iteration of HVP (Fig.~\ref{SM}e). The last
term is not reduced to HVP iteration and it is due to the hadronic
light-by-light (HLbL) scattering mechanism (Fig.~\ref{SM}g).

Hadronic contributions in (\ref{aStrong}) are determined by effects
dominated by long distance dynamics, the region where the methods of
perturbation theory of Quantum Chromodynamics (QCD) do not applicable and
one must use less reliable nonperturbative approaches. However, in case of
HVP, using analyticity and unitarity (the optical theorem) $a_{\mu }^{%
\mathrm{HVP}}$ can be expressed as the spectral representation integral \cite{BM61,Durand:1962zzb}
\begin{equation}
a_{\mu }^{\mathrm{HVP}}=\frac{\alpha }{\pi }\int_{4m_{\pi }^{2}}^{\infty }%
\frac{dt}{t}K(t)\rho _{\mathrm{V}}^{(\mathrm{H})}\left( t\right) ,
\label{Amm_rho}
\end{equation}%
which is a convolution of the hadronic spectral function
\begin{equation}
\rho _{V}^{\mathrm{(H)}}\left( t\right) =\frac{1}{\pi }\mathrm{Im}\Pi ^{\mathrm{(H)%
}}\left( t\right)  \label{rhoH}
\end{equation}%
with the known from QED kinematical factor%
\begin{equation}
K(t)=\int_{0}^{1}dx{\frac{x^{2}(1-x)}{x^{2}+(1-x)t/m_{\mu }^{2}}},
\label{Kfac}
\end{equation}%
where $m_{\mu }$ is the muon mass. The QED factor is sharply peaked at low
invariant masses $t$ and decreases monotonically with increasing $t$. Thus,
the integral defining $a_{\mu }^{\mathrm{HVP}}$ is sensitive
to the details of the spectral function $\rho _{V}^{\mathrm{(H)}}\left(
t\right) $ at low $t$. At present there is no direct theoretical tools that
allow to calculate the spectral function at low $t$ with required accuracy.
Fortunately, $\rho _{V}^{\mathrm{(H)}}\left( t\right) $ is related to the
total $e^{+}e^{-}\rightarrow \gamma ^{\ast }\rightarrow $ hadrons
cross-section $\sigma (t)$ at center-of-mass energy squared $t$ by
\begin{equation}
\sigma ^{e^{+}e^{-}\rightarrow \mathrm{hadrons}}(t)=\frac{4\pi \alpha }{t}%
\rho _{\mathrm{V}}^{(\mathrm{H})}\left( t\right) ,  \label{Sigma_rho}
\end{equation}%
and this fact is used to get quite accurate estimate of $a_{\mu }^{\mathrm{%
HVP}}$. The most precise recent phenomenological evaluations of $a_{\mu }^{%
\mathrm{HVP}}$, using recent $e^{+}e^{-}\rightarrow \mathrm{hadrons}$ data,
provide the results%
\begin{equation}
a_{\mu }^{\mathrm{HVP,LO,}e^{+}e^{-}}=\left\{
\begin{array}{l}
692.3~\left( 4.2\right) \times 10^{-10},\quad \text{\cite{Davier:2010nc}} \\
694.91~\left( 4.27\right) \times 10^{-10}.\quad \text{\cite{Hagiwara:2011af}}%
\end{array}%
\right.  \label{aHVPd}
\end{equation}%
In addition, data on inclusive decays of the $\tau $-lepton into hadrons are
used to replace the $e^{+}e^{-}$ data in certain energy regions. This is
possible, since the vector current conservation law relates the $I=1$ part
of the electromagnetic spectral function to the charged current vector
spectral function measured in $\tau \rightarrow \nu $ +non-strange hadrons (see, i.e. \cite%
{Jegerlehner:2011ti}).
All these allows to reach during the last decade a substantial improvement
in the accuracy of the contribution from the HVP.

Similar dispersion relation approach and the same phenomenological input
lead to the estimate of the next-to-leading hadronic contribution (Fig.~\ref%
{SM}e) \cite{Hagiwara:2011af,Kurz:2013exa}%
\begin{eqnarray}
a_{\mu }^{\mathrm{HVP,NLO}}=-9.84~\left( 0.07\right) \times 10^{-10},  \label{aHO}\\
a_{\mu }^{\mathrm{HVP,NNLO}}=1.24~\left( 0.01\right) \times 10^{-10}
\end{eqnarray}%
Thus, the HVP and next-to-leading order contribution related to HVP are known with an
accuracy better than 1 per cent.

In near future it is expected that new and precise measurements from CMD3
and SND at VEPP-2000 in Novosibirsk, BES III in Beijing and KLOE-2 at DAFNE
in Frascati allow to significantly increase accuracy of predictions for $%
a_{\mu }^{\mathrm{HVP}}$ and $a_{\mu }^{\mathrm{ho}}$ and resolve some
inconsistency problems between different set of data.

Subtracting from the experimental result the contributions from electroweak interaction and hadronic vacuum effect one gets
\begin{equation}
a_{\mu }^{\mathrm{BNL}}-a_{\mu }^{\mathrm{QED,Rb}}-a_{\mu }^{\mathrm{weak}}-a_{\mu }^{\mathrm{HVP}}=37.95~\left( 7.64\right) \times 10^{-10},
\label{DaEWHVP}
\end{equation}
where one can treat this number as an experimental result for the rest contributions, i.e. of the strong interaction of due to the light-by-light mechanism and of the hypothetical interactions beyond SM.

\section{Hadronic contributions to the muon $g-2$. Light-by-light scattering mechanism.}
\label{sec-4}

The basic element for calculations of the hadronic LbL contribution to the
muon AMM (Fig. \ref{SM}g) is the fourth-rank light quark hadronic vacuum polarization tensor
\begin{align}
&\mathrm{\Pi}_{\mu\nu\lambda\rho}(q_1,q_2,q_3)=\int d^4x_1\int d^4x_2\int
d^4x_3 e^{i(q_1x_1+q_2x_2+q_3x_3)}\times  \notag \\
&\quad\quad\quad\times\left<
0|T(j_\mu(x_1)j_\nu(x_2)j_\lambda(x_3)j_\rho(0))|0\right>,
\end{align}
where $j_\mu(x)$ are light quark electromagnetic currents and $\left|0\right>
$ is the QCD vacuum state.

The muon AMM can be extracted by using the projection \cite{Brodsky:1967sr}
\begin{align}
a_\mu^{\mathrm{LbL}}=\frac{1}{48m_\mu}\mathrm{Tr}\left((\hat{p}%
+m_\mu)[\gamma^\rho,\gamma^\sigma](\hat{p}+m_\mu)\mathrm{\Pi}%
_{\rho\sigma}(p,p)\right),  \notag
\end{align}
where
\begin{align}
&\mathrm{\Pi}_{\rho\sigma}(p^\prime,p)=-i e^6 \int \frac{d^4q_1}{(2\pi)^4}%
\int \frac{d^4q_2}{(2\pi)^4}\frac{1}{q_1^2 q_2^2 (q_1+q_2-k)^2}\times  \notag
\\
&\quad\quad\times \gamma^\mu \frac{\hat{p}^\prime-\hat{q}_1+m_\mu}{%
(p^\prime-q_1)^2-m_\mu^2}\gamma^\nu \frac{\hat{p}-\hat{q}_1-\hat{q}_2+m_\mu}{%
(p-q_1-q_2)^2-m_\mu^2} \gamma^\lambda \times  \notag \\
&\quad\quad \times \frac{\partial}{\partial k^\rho}\mathrm{\Pi}%
_{\mu\nu\lambda\sigma}(q_1,q_2,k-q_1-q_2), \quad
\end{align}
$m_\mu$ is the muon mass, $k_\mu=(p^\prime-p)_\mu$ and it is necessary to
consider the limit $k_\mu \to 0$.

\begin{figure*}[tbh]
\centerline{\includegraphics[width=12.7cm]{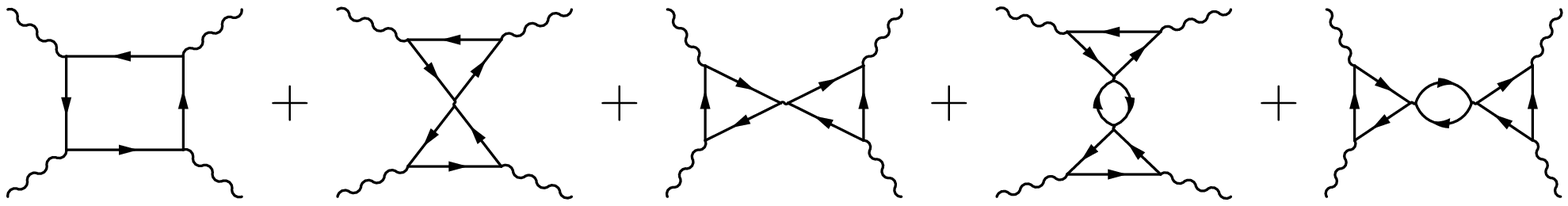}} \centerline{%
\includegraphics[width=7.62cm]{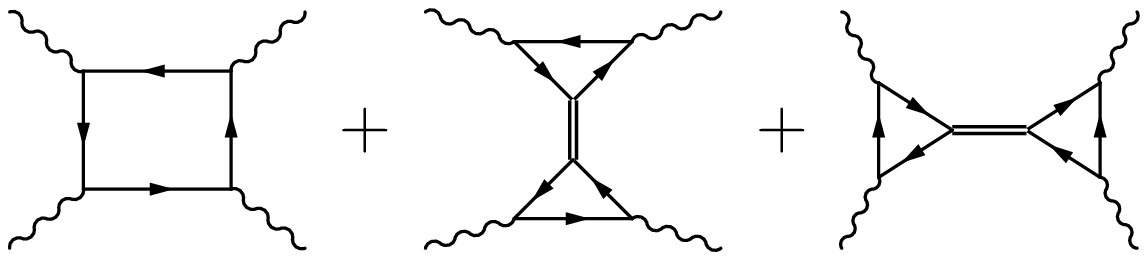}} \vspace*{8pt}
\caption{A schematic illustration for the diagrams contributing to the
four-rank polarization tensor to the leading in $1/N_{c}$ order. The four-fermion interaction is introduced through (\ref{33model}). The
nonlocal multi-photon vertices are not shown for simplicity, see Fig.
\protect\ref{Fig: BoxCont}.}
\label{BoxAll}
\end{figure*}

\begin{figure*}[t]
\centerline{\includegraphics[width=12.7cm]{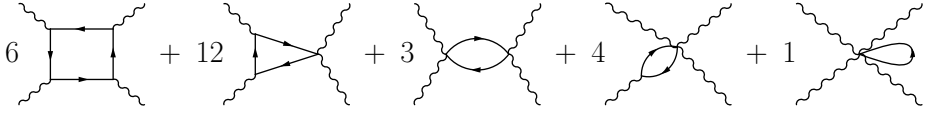}} \vspace*{8pt}
\caption{Contact terms which are gave contribution to $\mathrm{\Pi }_{%
\protect\mu \protect\nu \protect\lambda \protect\rho }(q_{1},q_{2},q_{3})$.
Numbers in front of diagrams are the degeneracy factors. }
\label{Fig: BoxCont}
\end{figure*}

In general, the HLbL scattering amplitude is a complicated object for
calculations. It is a sum of different diagrams including the dynamical
quark loop, the meson exchanges, the meson loops and the iterations of these
processes. Fortunately, already in the first papers devoted to the
calculation of the HLbL contributions \cite%
{deRafael:1993za,Hayakawa:1995ps,Bijnens:1995cc}, it has been recognized
that these numerous terms show a hierarchy. This is related to existence of
two small parameters: the inverse number of colors $1/N_{c}$ and the ratio
of the characteristic internal momentum to the chiral symmetry parameter $%
m_{\mu }/(4\pi f_{\pi })\sim 0.1$. The former suppresses the multiloop
contributions, so that the leading contribution is due to the quark loop
diagram and the two-loop diagrams with mesons in the intermediate state.
In latter case, the contribution of the diagram with intermediate pion is enhanced by
small pion mass in the meson propagator. The leading in $1/N_c$ diagrams are drawn in Fig. \ref{BoxAll}. They are the box diagram with dynamical quarks (Fig. \ref{Fig: BoxCont}) and the meson exchange diagrams in pseudoscalar, scalar and axial-vector channels.

For explicit calculations of the hadronic contributions to the muon AMM due to the light-by-light scattering mechanism \cite{Dorokhov:2008pw,Dorokhov:2011zf,Dorokhov:2012qa,Dorokhov:2014iva,Dorokhov:2015psa} we use
 the $SU(3)\times SU(3)$ chiral quark model with nonlocal Lagrangian (N$\chi$QM)
\begin{align}
\mathcal{L}& =\bar{q}(x)(i\hat{\partial}-m_{c})q(x)+\frac{G}{2}%
[J_{S}^{a}(x)J_{S}^{a}(x)+J_{PS}^{a}(x)J_{PS}^{a}(x)]  \notag \\
& -\frac{H}{4}%
T_{abc}[J_{S}^{a}(x)J_{S}^{b}(x)J_{S}^{c}(x)-3J_{S}^{a}(x)J_{PS}^{b}(x)J_{PS}^{c}(x)],
\label{33model}
\end{align}%
where $q\left( x\right) $ are the quark fields, $m_{c}$ $\left(
m_{u}=m_{d}\neq m_{s}\right) $ is the diagonal matrix of the quark current
masses, $G$ and $H$ are the four- and six-quark coupling constants. Second
line in the Lagrangian represents the Kobayashi--Maskawa--t`Hooft
determinant vertex with the structural constant
\begin{equation*}
T_{abc}=\frac{1}{6}\epsilon _{ijk}\epsilon _{mnl}(\lambda _{a})_{im}(\lambda
_{b})_{jn}(\lambda _{c})_{kl},
\end{equation*}%
where $\lambda _{a}$ are the Gell-Mann matrices for $a=1,..,8$ and $\lambda
_{0}=\sqrt{2/3}I$.
The nonlocal structure of the model is introduced via the nonlocal quark
currents
\begin{equation}
J_{M}^{a}(x)=\int d^{4}x_{1}d^{4}x_{2}\,f(x_{1})f(x_{2})\,\bar{Q}%
(x-x_{1},x)\,\Gamma _{M}^{a}Q(x,x+x_{2}),  \label{JaM}
\end{equation}%
where $M=S$ for the scalar and $M=PS$ for the pseudoscalar channels, $\Gamma
_{{S}}^{a}=\lambda ^{a}$, $\Gamma _{{PS}}^{a}=i\gamma ^{5}\lambda ^{a}$ and $%
f(x)$ is a form factor with the nonlocality parameter $\Lambda $ reflecting
the nonlocal properties of the QCD vacuum.
In (\ref{JaM}), the gauge-invariant interaction with an external photon field $V_{\mu }^{a}$
is introduced through the Schwinger phase factor%
\begin{equation}
Q\left( x,y\right) =\mathcal{P}\exp \left\{
i\int_{x}^{y}dz^{\mu }V_{\mu }^{a}\left( z\right) T^{a}\right\} q\left(
y\right) .  \label{SchwPhF}
\end{equation}%
In order to guarantee the Ward-Takahashi identities, it induces the quark-antiquark--$n$-photon vertices. Additionally, there
appear the meson--quark-anti-quark--$n$-photon vertices.

\begin{figure*}[t]
\begin{center}
\centerline{\includegraphics[width=0.45\textwidth]{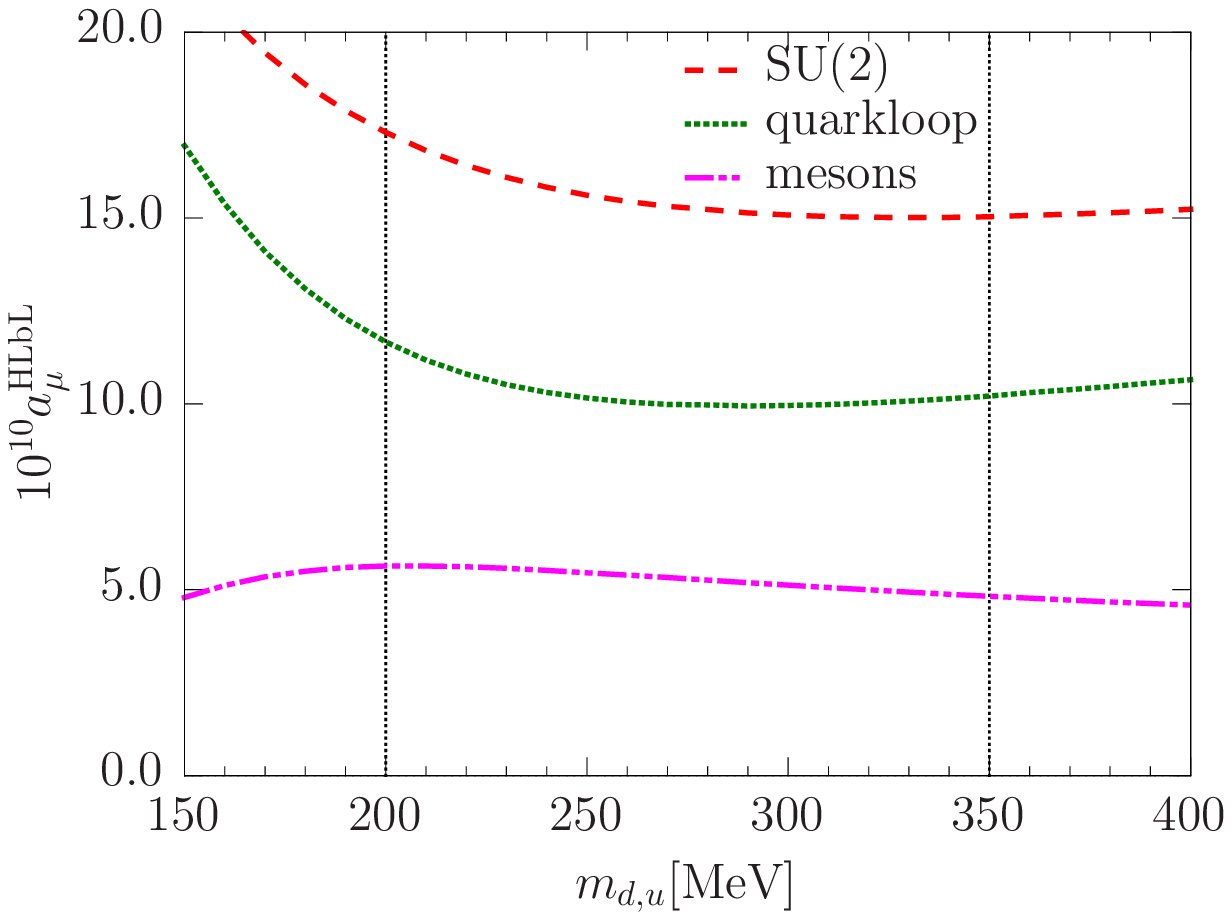} 
\includegraphics[width=0.45\textwidth]{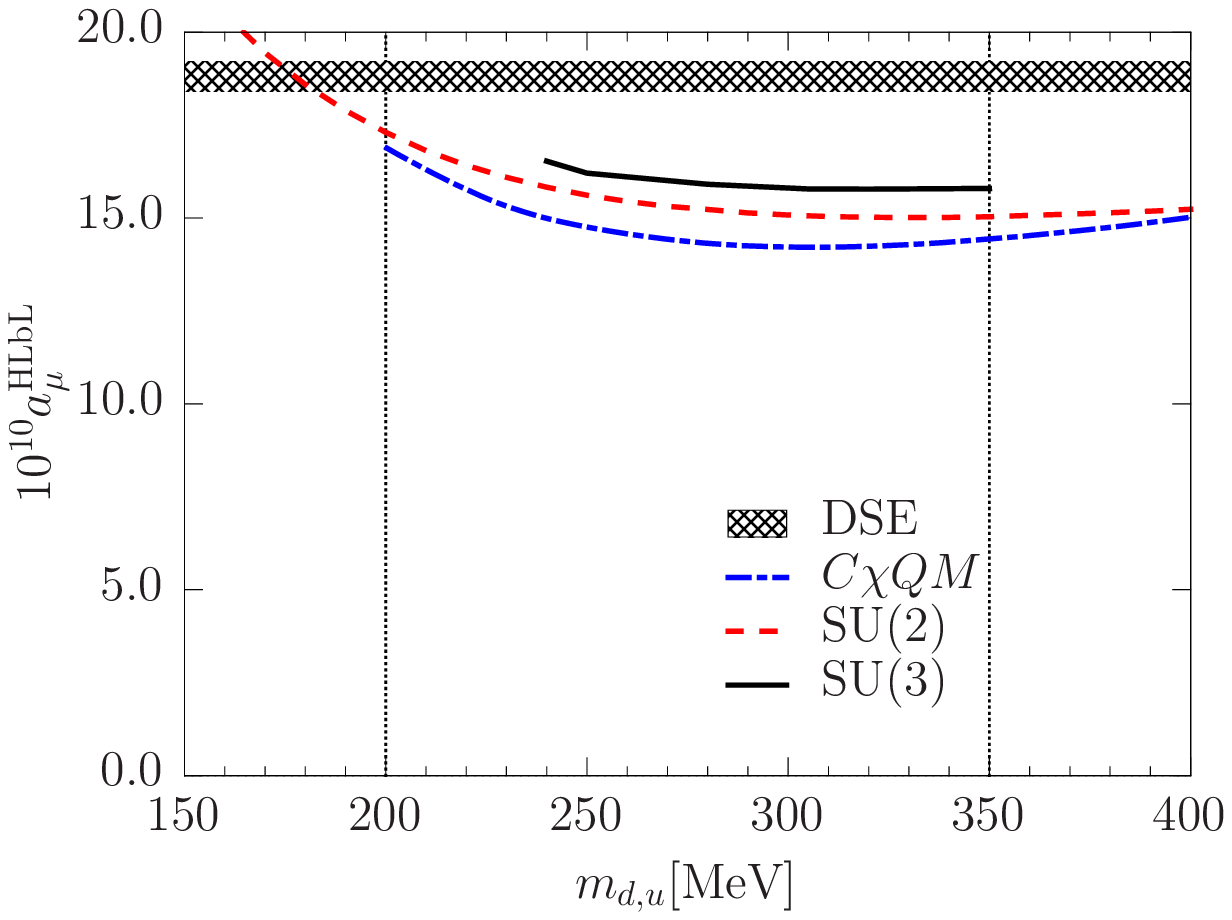}} 
\end{center}
\caption{(Left) The results for $a_{\mu}^{\mathrm{HLbL}}$ in the $SU(2)$ model: the
red dashed line is the total result, the green dotted line is the quark loop
contribution and the magenta dash-dot-dot line is the $\pi+\sigma$ contribution. Thin
vertical line indicates the region for estimation of $a_{\mu}^{\mathrm{HLbL}}$ error band.
(Right) The results for $a_{\mu}^{\mathrm{HLbL}}$: the black solid line is the $SU(3)$-result,
the red dash line corresponds to the $SU(2)$-result,
the blue dash-dotted line is the $C\chi QM$ result \cite{Greynat:2012ww},
hatched region correspond to DSE result \cite{Fischer:2010iz}.}
\label{fig: 16}%
\end{figure*}

The numerical results for the value of $a_{\mu}^{\mathrm{HLbL}}$ are presented
in Fig. \ref{fig: 16} for the $SU(2)$ and $SU(3)$ models.
The estimates for the partial contributions to $a_{\mu}^{\mathrm{HLbL}}$ (in
$10^{-10}$) are the $\pi^{0}$ contribution $5.01(0.37)$ \cite{Dorokhov:2011zf}%
, the sum of the contributions from $\pi^{0}$, $\eta$ and $\eta^{\prime}$
mesons $5.85(0.87)$ \cite{Dorokhov:2011zf}, the scalar $\sigma$, $a_{0}(980)$
and $f_{0}(980)$ mesons contribution $0.34(0.48)$
\cite{Dorokhov:2014iva,Dorokhov:2012qa}, and the quark loop contribution is
$11.0(0.9)$ \cite{Dorokhov:2014iva}. The total contribution obtained in the
leading order in the $1/N_{c}$ expansion is (see also \cite{Dorokhov:2014iva}%
)
\begin{equation}
a_{\mu}^{\mathrm{HLbL,N\chi QM}}=16.8(1.25)\cdot10^{-10}. \label{aNchiQM}%
\end{equation}
The error bar accounts for the spread of the results depending on the model
parameterizations. Comparing with other model calculations, we conclude that
our results are quite close to the recent results obtained in
\cite{Fischer:2010iz,Greynat:2012ww}.

If we add the result (\ref{aNchiQM}) to all other known contributions of the
standard model to $a_{\mu}$, (\ref{aMuQED}),(\ref{aHVPd}) and (\ref{aHO}),
we get that the difference between experiment (\ref{amuBNL2}) and theory is
\begin{equation}
a_{\mu}^{\mathrm{BNL,CODATA}}-a_{\mu}^{\mathrm{SM}}=18.73\times10^{-10},
\end{equation}
which corresponds to $2.43\sigma$. If one uses the hadronic vacuum
polarization contribution from the $\tau$ hadronic decays instead of
$e^{+}e^{-}$ data \cite{Davier:2010nc}
\begin{equation}
a_{\mu}^{\mathrm{HVP,LO-\tau}}=701.5(4.7)\times10^{-10},
\end{equation}
the difference decreases  to $12.14\times10^{-10}$ ($1.53\sigma$) in the N$\chi$QM model (\ref{aNchiQM}).

\section{Conclusions}
\label{sect-4}

Clearly, a further reduction of both the experimental and theoretical
uncertainties is necessary. On the theoretical side, the calculation of the
still badly known hadronic light-by-light contributions in the next-to-leading
order in the $1/N_{c}$ expansion 
(the pion and kaon loops) and extension of the model by including heavier vector and axial-vector mesons
is the next goal.
The contribution of these effects and the model error induced by them are not included in the result (\ref{aNchiQM}).
Preliminary studies \cite{Hayakawa:1995ps,Melnikov:2003xd} show that these contributions
are one order smaller than the pseudoscalar exchanges and the quark loop contributions.
However, the interesting point that inclusion of vector
channel can strongly suppress contribution from the quark loop
due photon--vector meson exchange which lead
to appearance in each photon vertex additional VMD-like factor.
This was found in local NJL model \cite{Bijnens:1995cc} and should be
carefully investigated in the nonlocal one.

New experiments at FNAL and J-PARC have to resolve the muon $g-2$ problem, increasing effect or leading to its disappearance.

Numerical calculations are performed on computing cluster "Academician V. M. Matrosov" (http://hpc.icc.ru). The work is supported by Russian Science Foundation grant (RSCF 15-12-10009).


\begin{thebibliography}{34}

\bibitem{Kush:1947}
P.~Kush, H.M. Foley, Phys.Rev. \textbf{72}, 1256 (1947)

\bibitem{Kush:1948}
H.M. Foley, P.~Kush, Phys.Rev. \textbf{73}, 412 (1948)

\bibitem{Schwinger:1948iu}
J.S. Schwinger, Phys.Rev. \textbf{73}, 416 (1948)

\bibitem{Hanneke:2008tm}
D.~Hanneke, S.~Fogwell, G.~Gabrielse, Phys. Rev. Lett. \textbf{100}, 120801
  (2008), \texttt{0801.1134}

\bibitem{Laporta:1996mq}
S.~Laporta, E.~Remiddi, Phys. Lett. \textbf{B379}, 283 (1996),
  \texttt{hep-ph/9602417}

\bibitem{Aoyama:2012wj}
T.~Aoyama, M.~Hayakawa, T.~Kinoshita, M.~Nio, Phys. Rev. Lett. \textbf{109},
  111807 (2012), \texttt{1205.5368}

\bibitem{KINOSHITA:2014uza}
T.~Kinoshita, Int. J. Mod. Phys. \textbf{A29}, 1430003 (2014)

\bibitem{Bouchendira:2010es}
R.~Bouchendira, P.~Clade, S.~Guellati-Khelifa, F.~Nez, F.~Biraben, Phys. Rev.
  Lett. \textbf{106}, 080801 (2011), \texttt{1012.3627}

\bibitem{Bennett:2006fi}
G.~Bennett, others (Muon~g 2~collaboration) (Muon g-2), Phys.Rev. \textbf{D73},
  072003 (2006), \texttt{hep-ex/0602035}

\bibitem{Mohr:2012tt}
P.J. Mohr, B.N. Taylor, D.B. Newell, Rev. Mod. Phys. \textbf{84}, 1527 (2012),
  \texttt{1203.5425}

\bibitem{Agashe:2014kda}
K.A. Olive et~al. (Particle Data Group), Chin. Phys. \textbf{C38}, 090001
  (2014)

\bibitem{Venanzoni:2012sq}
G.~Venanzoni, J. Phys. Conf. Ser. \textbf{349}, 012008 (2012)

\bibitem{Saito:2012zz}
N.~Saito (J-PARC g-'2/EDM), AIP Conf. Proc. \textbf{1467}, 45 (2012)

\bibitem{Aoyama:2012wk}
T.~Aoyama, M.~Hayakawa, T.~Kinoshita, M.~Nio, Phys. Rev. Lett. \textbf{109},
  111808 (2012), \texttt{1205.5370}

\bibitem{Czarnecki:2002nt}
A.~Czarnecki, W.J. Marciano, A.~Vainshtein, Phys. Rev. \textbf{D67}, 073006
  (2003), [Erratum: Phys. Rev.D73,119901(2006)], \texttt{hep-ph/0212229}

\bibitem{Gnendiger:2013pva}
C.~Gnendiger, D.~St{\"o}ckinger, H.~St{\"o}ckinger-Kim, Phys. Rev. \textbf{D88},
  053005 (2013), \texttt{1306.5546}

\bibitem{BM61}
C.~Bouchiat, L.~Michel, J. Phys. Radium \textbf{22}, 121 (1961)

\bibitem{Durand:1962zzb}
L.~Durand, Phys. Rev. \textbf{128}, 441 (1962)

\bibitem{Davier:2010nc}
M.~Davier, A.~Hoecker, B.~Malaescu, Z.~Zhang, Eur.Phys.J. \textbf{C71}, 1515
  (2011), \texttt{1010.4180}

\bibitem{Hagiwara:2011af}
K.~Hagiwara, R.~Liao, A.D. Martin, D.~Nomura, T.~Teubner, J. Phys.
  \textbf{G38}, 085003 (2011), \texttt{1105.3149}

\bibitem{Jegerlehner:2011ti}
F.~Jegerlehner, R.~Szafron, Eur. Phys. J. \textbf{C71}, 1632 (2011),
  \texttt{1101.2872}

\bibitem{Kurz:2013exa}
A.~Kurz, T.~Liu, P.~Marquard, M.~Steinhauser, Nucl. Phys. \textbf{B879}, 1
  (2014), \texttt{1311.2471}

\bibitem{Brodsky:1967sr}
S.J. Brodsky, E.~De~Rafael, Phys. Rev. \textbf{168}, 1620 (1968)

\bibitem{deRafael:1993za}
E.~de~Rafael, Phys.Lett. \textbf{B322}, 239 (1994), \texttt{hep-ph/9311316}

\bibitem{Hayakawa:1995ps}
M.~Hayakawa, T.~Kinoshita, A.I. Sanda, Phys. Rev. Lett. \textbf{75}, 790
  (1995), \texttt{hep-ph/9503463}

\bibitem{Bijnens:1995cc}
J.~Bijnens, E.~Pallante, J.~Prades, Phys. Rev. Lett. \textbf{75}, 1447 (1995),
  [Erratum: Phys. Rev. Lett.75,3781(1995)], \texttt{hep-ph/9505251}

\bibitem{Dorokhov:2008pw}
A.E. Dorokhov, W.~Broniowski, Phys. Rev. \textbf{D78}, 073011 (2008),
  \texttt{0805.0760}

\bibitem{Dorokhov:2011zf}
A.E. Dorokhov, A.E. Radzhabov, A.S. Zhevlakov, Eur. Phys. J. \textbf{C71}, 1702
  (2011), \texttt{1103.2042}

\bibitem{Dorokhov:2012qa}
A.E. Dorokhov, A.~Radzhabov, A.S. Zhevlakov, Eur.Phys.J. \textbf{C72}, 2227
  (2012), \texttt{1204.3729}

\bibitem{Dorokhov:2014iva}
A.E. Dorokhov, A.E. Radzhabov, A.S. Zhevlakov, JETP Lett. \textbf{100}, 133
  (2014), \texttt{1406.1019}

\bibitem{Dorokhov:2015psa}
A.E. Dorokhov, A.E. Radzhabov, A.S. Zhevlakov, Eur. Phys. J. \textbf{C75}, 417
  (2015), \texttt{1502.04487}

\bibitem{Greynat:2012ww}
D.~Greynat, E.~de~Rafael, JHEP \textbf{07}, 020 (2012), \texttt{1204.3029}

\bibitem{Fischer:2010iz}
C.S. Fischer, T.~Goecke, R.~Williams, Eur. Phys. J. \textbf{A47}, 28 (2011),
  \texttt{1009.5297}

\bibitem{Melnikov:2003xd}
K.~Melnikov, A.~Vainshtein, Phys. Rev. \textbf{D70}, 113006 (2004),
  \texttt{hep-ph/0312226}

\end{thebibliography}

\end{document}